\documentclass[12pt]{article}
\usepackage[super,sort&compress]{natbib}

\usepackage{setspace}
\usepackage[margin = 3cm]{geometry}

\usepackage[french, english]{babel}
\usepackage[T1]{fontenc}
\usepackage{tikz}
\usepackage{helvet}
\usepackage{multirow}
\usepackage{amsmath}
\usepackage{amssymb}

\usepackage[figuresright]{rotating}
\usepackage{bbm}

\usepackage[colorlinks,bookmarksopen,bookmarksnumbered,citecolor=red,urlcolor=red]{hyperref}

\newcommand{\RN}[1]{
  \textup{\uppercase\expandafter{\romannumeral#1}}%
}

\def\bSig\mathbf{\Sigma}

\usepackage{subfig}
\usepackage{float}
\usepackage{graphics}
\usepackage{graphicx}
\usepackage{xcolor}

\title{Including an infrequently measured time-dependent error-prone covariate in survival analyses: a simulation-based comparison of methods}

\begin{document}

\begin{titlepage}
\author{Viviane Philipps$^1$, Laurence Freedman$^2$ , Veronika Deffner$^3$, Catherine Helmer$^1$,  \\
Hendriek Boshuizen$^4$,
Anne C.M. Thi\'ebaut$^5$, C\'ecile Proust-Lima$^{1*}$,\\
on behalf of Measurement Error and Misclassification \\Topic Group (TG4) of the STRATOS Initiative}

\maketitle

\noindent
$^1$ Univ. Bordeaux, Inserm, Bordeaux Population Health Research Center, F-33000 Bordeaux, France\\
$^2$ The Gertner Institute of Epidemiology and Health Policy Research, Sheba Medical Center, Tel Hashomer, Israel\\
$^3$ Statistical Consulting Unit StaBLab, Department of Statistics, Ludwig-Maximilians-Universität, Munich 80539, Germany\\
$^4$ Division of Human Nutrition and Health, Wageningen University \& Research, 6708 WE Wageningen, The Netherlands\\
$^5$ Université Paris-Saclay, UVSQ, Inserm, CESP, High Dimensional Biostatistics Team, 94807 Villejuif, France\\
$^*$ {\bf Correspondence to:} cecile.proust-lima@inserm.fr
\end{titlepage}

{\bf Abstract}
Epidemiologic studies often evaluate the association between an exposure and an event risk. When time-varying, exposure updates usually occur at discrete visits although changes are in continuous time and survival models require values to be constantly known. Moreover, exposures are likely measured with error, and their observation truncated at the event time. 
We aimed to quantify in a Cox regression the bias in the association resulting from intermittent measurements of an error-prone exposure. Using simulations under various scenarios, we compared five methods: last observation carried-forward (LOCF), classical two-stage regression-calibration using measurements up to the event (RC) or also after (PE-RC), multiple imputation (MI) and joint modeling of the exposure and the event (JM). 
The LOCF, and to a lesser extent the classical RC, showed substantial bias in almost all 43 scenarios. The RC bias was avoided when considering post-event information. The MI performed relatively well, as did the JM. Illustrations exploring the association of Body Mass Index and Executive Functioning with dementia risk showed consistent conclusions. 
Accounting for measurement error and discrete updates is critical when studying time-varying exposures. MI and JM techniques may be applied in this context, while classical RC should be avoided due to the informative truncation.\\

\textbf{Keywords:} time-varying exposure, survival model, classical measurement error, longitudinal data, simulation study\\

\newpage

\section*{INTRODUCTION}
The association between a time-varying exposure and a time-to-event outcome is often under investigation in epidemiological research. Proportional hazards models, including the semi-parametric Cox model, are by far the most favored models to study such associations \citep{cox1972}.
However, in addition to the assumptions of proportional hazards over time and log-linearity of the relationship between the hazard and the exposure, the Cox model relies on three stringent assumptions that are rarely met when the exposure varies over time. First, the individual exposure is supposed to be continuously recorded so that it is known at any time during the observation period. Beyond age or time since treatment initiation, this may for instance occur in intensive care where biomarkers are measured continuously, or in environmental epidemiology with continuous measurements of air pollution. However, in many situations, the design involves staggered visits and the exposure information becomes infrequently measured and intermittently unknown. Second, as most statistical models, the Cox model assumes that the exposure is measured without error. Yet, every measurement is subject to error, either due to the measuring device or to the impossibility to access precisely the actual construct. When dealing with human-related processes, this error may become substantial and cannot be ignored anymore \cite{prentice1982}. Third, the Cox model assumes that the exposure is exogenous, that is the future exposure process (i.e. after the event occurs) is independent of the event occurrence \cite{kalbfleisch}. This might be true for environmental exposures, such as air pollution, 
but it rarely applies beyond that. Most often, exposures are endogenous with the post-event values likely impacted by the event; for instance, body mass index or other lifestyle behaviors after dementia onset. 

In the presence of a time-varying covariate that is infrequently measured, the most common, but naive, strategy of analysis is to carry forward the last observation until a new one becomes available. This method, also known as LOCF or LVCF (Last Observation/Value Carried Forward), tends to propagate the measurement errors in both the timings and the exposure values, as demonstrated in simulation studies \citep{arisido,liao,tsiatis,wang}. 
Andersen and Liest\o l \cite{andersen} have acknowledged a resulting attenuation of the regression coefficient when the covariate is measured at infrequent times, and proposed a correction.

When the error-prone exposure is measured over time, a solution to retrieve the error-free underlying exposure process is to model its trajectory over time. This is what linear mixed models aim to do by isolating the measurement error and parametrically describing the remaining individual signal \cite{laird1982}. The predicted individual exposure can be calculated at any time and may then serve as a proxy of the error-free exposure in the Cox model \cite{ye2008}. This procedure follows the regression calibration (RC) approach largely recommended in the presence of measurement error \cite{shaw2018}. The individual prediction in RC approximates the covariate with a Berkson error, which, in contrast with classical error, does not induce bias in the estimation of the regression coefficients in linear models \cite{boe}. The longitudinal design raises however an additional issue. When the time-dependent exposure is endogenous, its collection ends when the event occurs, leading to a truncation of the exposure process which is likely informative \cite{geskus}. For instance, if the exposure is a risk factor for the event, participants with worse trajectories are likely to experience the event earlier and thus have shorter follow-ups.
Several simulation studies have demonstrated that regression calibration for time-dependent exposure in survival models leads to biased results \cite{liao,tsiatis,wang,ye2008} due to the early truncation of the exposure collection linked to the event which breaks down the RC procedure \cite{boe}. 

A workaround is to estimate a linear mixed model at each time point where the covariate's value is needed (each failure time), using only the data available up to that time point. This method, introduced in the 1990s as the sequential two-stage approach and also known as risk-set regression calibration \cite{liao} is however numerically very demanding. In the late 1990s, joint models of the exposure process and the hazard function \cite{rizopoulosbook,wulfsohn} have also emerged. By simultaneously employing a mixed model for retrieving the underlying error-free exposure and a survival model for associating it with the event rate over time, they naturally account for the infrequent measures, error-prone observations, and the internal nature of the exposure. Joint models have gained significant popularity and are now widely adopted within the statistical community. However, despite efforts to make software user-friendly \cite{rustand,hickey, jmbayes2}, the joint modeling approach remains a sophisticated method that needs advanced statistical expertise and is computationally demanding. This may explain why LOCF methods and classical regression calibration are still widely used in epidemiology. Facing the complexity of joint model estimation, particularly in the context of multiple time-dependent exposures, a two-stage procedure relying on the multiple imputation principle was also proposed \cite{moreno}.

The international STRATOS initiative (http://stratos-initiative.org) seeks to Strengthening the Analytic Thinking in the design and analysis of Observational Studies. As part of STRATOS topic group TG4 dedicated to "measurement error and misclassification", we aimed to address the issue of infrequent and error-prone measurement of time-varying exposures in epidemiological survival analyses. This work reviews five statistical approaches available and compares them in simulations and real data analyses. We conclude with recommendations and cautions regarding appropriate and easy-to-implement methods.

\section*{METHODOLOGY} \label{sec:methods}

Let $X(t)$ be the time-varying exposure process we are interested in and whose association with an event is to be assessed. The right-censored observed  event time is denoted $T_i$ for individual $i$ ($i=1,...,N$) with event indicator $E_i$. We assume that the process $X_i(t)$ defined at any time $t$ is not known for all the individuals. We only observe a noisy version of it, $X^*_{ij}$, subject to homoskedastic classical measurement error at study visits $j$ ($j=1,n_i$) so that: 
$$X^*_{ij} = X^*_i(t_{ij}) = X_i(t_{ij}) + \epsilon_{ij}$$

\noindent
where $\epsilon_{ij}$ is independent of $X_i(t_{ij})$ and has mean zero and constant variance. 
 
We assume that the target model of interest (considered as well-specified) for studying the association of the time to event and the exposure is following a proportional hazards model:
\begin{equation} \label{targetM}
\lambda_i(t) = \lambda_0(t)\exp(X_i(t)\gamma),
\end{equation}

\noindent where $\lambda_0(t)$ is the baseline hazard function (left unspecified in the Cox model) and $\gamma$ quantifies the strength of association between the exposure and the instantaneous risk of event, both considered at the exact same time $t$. No adjustment covariates were added here for sake of readability. Since $X$ is not observed, this model cannot be directly estimated and special estimation methods must be used. They are summarized in Figure \ref{fig:methods} and further detailed below. 

\subsection*{The LOCF Cox model}\label{subsec:naive}

The most popular solution is to replace $X_i(t)$ by its noisy version that was observed at the last visit preceding $t$. This approach thus assumes that, after each measurement, $X^*$ remains constant until a new measurement becomes available. This technique, made popular under the name LOCF for last observation carried forward, can be defined as follows. The LOCF-approximated process is $X^{*\text{LOCF}}_i(t)=X^*_i(t_{ij})$ for all $t \in [t_{ij}, t_{ij+1})$ ($j=1,...,n_i$) with $t_{in_i+1}=T_i$, and the associated Cox model is:

\begin{equation}
\lambda_i(t) = \lambda_0(t)\exp(X^{*\text{LOCF}}_i(t)~ \gamma^\text{LOCF})
\end{equation}

This method is very easy to implement in any standard software. However, it is important to note that this model not only relies on a strong and unlikely piecewise constant value, but also neglects the measurement error in $X^*$.

\subsection*{Cox regression calibration} \label{subsec:RC}

Regression calibration is a two-stage procedure recommended in the presence of measurement error \cite{shaw2018,boe}. It involves generating a prediction for $X$ (which will have Berkson error rather than classical error \cite{boe}) and subsequently including this prediction into the model of interest. In our context, we obtain a predicted value of $X(t)$ from a linear mixed calibration model:
\begin{equation}\label{RCMM}
  \begin{aligned}
    X^*_{ij} &= X_i(t_{ij}) + \varepsilon_{ij} \\ & =  F(t_{ij})^\top \beta + F(t_{ij})^\top u_i + \varepsilon_{ij}  \\
    \end{aligned}
  \end{equation}

\noindent where $F(t_{ij})$ is a pre-determined vector of time functions that can approximate the general shape of $X^*_i$ trajectories  over time, $\beta$ are the associated parameters at the population level, $u_i \sim \mathcal{N}(0, B)$ are the associated random effects at the individual level, and $\varepsilon_{ij} \sim \mathcal{N}(0, \sigma^2)$ are independent homoskedastic measurement errors. 
Once the model is estimated, the individual exposure process $X_i(t)$ can be predicted at any time $t$: $X^\text{RC}_i(t) = F(t_{ij})^\top \hat{\beta} + F(t_{ij})^\top \hat{u}_i$ with $\hat{\beta}$ the fixed-effect estimates and $\hat{u}_i$ the predicted random-effects. 

The second stage consists in estimating the Cox model where $X_i(t)$ is replaced by its prediction $X^\text{RC}_i(t)$:

\begin{equation}
\lambda_i(t) = \lambda_0(t)\exp(X^\text{RC}_i(t) ~ \gamma^\text{RC})
\end{equation}

This method handles the infrequent timing of the exposure measurements, as well as the measurement error. However, in most cases, the observation of $X^*(t)$ is truncated when the event occurs (Figure \ref{fig:methods} panel B) which can induce an informative dropout mechanism to which the mixed model is not robust, leading to biased parameter estimates. In the very specific situations where the exposure is exogenous, and still observable and meaningful after the event (e.g., air pollution), the mixed model estimates based on the complete exposure process (Figure \ref{fig:methods} panel C) do not suffer from this informative dropout.


The classical regression calibration technique can be easily implemented in any standard software using a mixed model estimation procedure and then a Cox model estimation procedure. In this work, we used \texttt{hlme} and \texttt{coxph} from the \texttt{lcmm} and \texttt{survival} R packages, respectively. However, the variance estimation of $\hat{\gamma}^\text{RC}$ needs attention. The standard variance estimates reported in the Cox model are not valid as they do not account for the first stage estimation uncertainty of the calibrated exposure $X^\text{RC}_i(t)$. A parametric bootstrap technique with 500 replicates was used to compute the total variance \citep{efron1994} (see section 1 of the supplementary material for details). 

In this work, we only investigated the classical regression calibration technique, not the risk-set regression calibration \cite{liao} for which no standard software was available. 

\subsection*{The imputation-based Cox model}\label{subsec:MI}

As measurement error, sparsity of measurement times and early truncation lead to $X(t)$ being unobserved, the analysis could be considered to have a missing data problem, and a multiple imputation technique can be used. Similar to the regression calibration technique, this is a two-stage method. However, there are two key differences: 

\begin{enumerate}
    \item The linear mixed model estimated in the first stage now includes information about the event:
\begin{equation}
  \begin{aligned}
    X^*_{ij} &= X(t_{ij}) + \varepsilon_{ij} \\ & =  F(t_{ij})^\top \beta + F(t_{ij})^\top u_i + G(t_{ij})^\top \zeta + \varepsilon_{ij} \\ 
    \end{aligned}
\end{equation}
with $G(t_{ij})$ a vector of covariates describing the event risk and $\zeta$ the corresponding regression coefficients. Following Moreno-Betancur et al \cite{moreno}, we included in $G(t_{ij})$ both the Nelson-Aalen estimator of the cumulative hazard at event time $\hat{\Lambda}_i(T_i)$ and the indicator $\mathbb{I}_{t_{ij}=t_{in_i}}$ which indicates the last measurement time prior to the event. Inclusion of these variables handles the problem of informative dropout.

    \item The underlying exposure process $X(t)$ is predicted based on samples from the first stage model: ${X}^\text{MI}_i(t) = F(t)(\tilde{\beta} + \tilde{u}_i) + G(t)\tilde{\zeta}$ where $\tilde{u}_i$ is a random draw sampled from the posterior distribution of the random effects for subject $i$. The fixed effects $\tilde{\beta}$ and $\tilde{\zeta}$, and the variance of the random effects are also sampled to further account for the uncertainty in the first-stage model parameters. 
\end{enumerate}


The second stage then consists in estimating the Cox model where $X_i(t)$ is replaced by its random draw ${X}^\text{MI}_i(t)$:
\begin{equation}
\lambda_i(t) = \lambda_0(t)\exp(X^\text{MI}_i(t) ~ \gamma^\text{MI})
\end{equation}

The final estimate of $\gamma^\text{MI}$ is the average estimate across the imputation replicates. The total variance of $\gamma^\text{MI}$ is computed using a parametric Bootstrap principle (as for the regression calibration, see supplementary Section 1) and the Rubin's rule \cite{rubin}. 


\subsection*{Joint model}\label{subsec:JM}

In contrast to the regression calibration technique and the imputation-based Cox model, the joint model simultaneously estimates the exposure trajectory and the survival model of interest \citep{rizopoulosbook}. As such, it inherently handles their interdependence, the error measurement, the sparse measurement times and the informative dropout. With the same notations and specifications as before, the model can be defined as follows:
\begin{equation}
    \left \{ 
    \begin{array}{ll}
    X^*_{ij} &= X_i(t_{ij}) + \varepsilon_{ij} =  F(t_{ij})^\top \beta + F(t_{ij})^\top u_i + \varepsilon_{ij}  \\
    \lambda_i(t) &= \lambda_0(t)\exp(X_i(t)\gamma^\text{JM}) \end{array} \right . \label{JMformulas}
\end{equation}


The joint model yields an estimate of $\gamma$ and a valid standard error. Despite user-friendly software solutions (e.g., JM \cite{rizopoulosbook}, joineRML \cite{hickey}, INLAjoint \cite{rustand} or JMbayes2 \cite{jmbayes2} in R), the joint model estimation is more complicated than the others and requires special attention regarding the specification of the estimation algorithm and convergence. The R package \texttt{JM} combined with a reliable optimization algorithm \cite{philipps} was used in this work. The variance of the parameters was estimated by the inverse of the Hessian. 

\section*{NUMERICAL SIMULATIONS}


\subsection*{Aim}
We conducted a simulation study to assess the performance of the LOCF Cox, Cox regression calibration, the imputation-based Cox model and the joint model to correctly estimate the association parameter $\gamma$ of the time-varying exposure measured at irregular times with classical measurement error, under a variety of scenarios exploring the magnitude of association with the event, the rate of truncation by the event, the magnitude of exposure measurement error and the shape of the exposure trajectory. The reporting follows guidelines for conducting and reporting simulation studies \cite{morris}. 


\subsection*{Data generation}
The data generation procedure is fully described in the supplementary Section 2 and Figure S1. For all 43 scenarios, the data were generated according to the target joint model (equation \eqref{JMformulas}).  
We defined linear exposure trajectories with mutually independent random intercept and random slope. The frequency of visits was every two years for a maximum of 10 years, and the exposure measurement errors had either smaller or larger variance compared to the true exposure variance (see Figure S2). In the smaller error setting, the error variance was equal to the true exposure variance at baseline and represented about 11\% of the exposure variance by time t=4. For the larger error setting, the error variance was 9 times more important than the exposure error at baseline and these variances became equal by time t=4. 
We generated three baseline hazard functions to investigate different frequencies of events and censoring, and considered two magnitudes of association between the exposure and the risk of event (labeled as weaker and stronger). These combinations resulted in 12 main scenarios (scenarios 1-12). Additional scenarios considered correlated random effects, quadratic trajectories, and intermittent missing visits. See Supplementary Table {S1} for a summary of all the scenarios.

\subsection*{Estimand}
The estimand of interest was the association coefficient $\gamma$ which quantifies the association between the time-varying exposure level and the instantaneous risk of event at the exact same time.

\subsection*{Models compared}
We generated 500 samples of $N=500$ subjects. The four methods (LOCF Cox, Cox regression calibration (RC), Cox model based on multiple imputation (MI) and joint model) were applied after truncation of the exposure observation at the time of event. For Cox regression calibration, we also considered the case where the exposure could still be observed after the event until the maximum of 10 years (denoted as post-event RC or PE-RC). For all the techniques (except LOCF), the same shape of exposure trajectory and the same association with the event as in the generation model were modeled. A replication script can be found \href{https://github.com/VivianePhilipps/Paper\_TimeVar\_STRATOS}{here}. For Cox RC and the MI Cox, the total variance was computed using 500 bootstrap vectors. 

\subsection*{Performance}
The performance measures were the bias shown with the empirical distribution of the association estimate $\gamma$, the empirical coverage rate of its 95\% confidence interval, and the Mean Squared Error (MSE). 

\subsection*{Results}
Figures \ref{fig:sim1} and \ref{fig:sim2} report the results of the 12 main scenarios under weaker ($\gamma=0.2$) and stronger ($\gamma=0.4$) association, respectively. 
{Under weaker association, smaller magnitude of measurement error and medium to high baseline survival (scenarios 1-2), the LOCF approach showed negligible bias. However, as soon as event time censoring became substantial, i.e. with low baseline survival (scenarios 3, 6, 9, and 12), or as the magnitude of measurement error increased (scenarios 4-6, 10-12), or the association became stronger (scenarios 7-12), the LOCF approach became substantially biased with effect attenuation up to 81\% of the true value in our scenarios. The regression calibration technique exhibited biases in the same scenarios although with less attenuation (up to 40\%). This bias was largely avoided when the post-event information was included, thereby eliminating the early informative truncation by the event (post-event RC), except in scenarios 10-12 with strong association and large measurement errors. 
The MI Cox model performed relatively well despite a small amount of bias (around 5\%) in scenarios with large measurement error (4-6,10-12). This resulted in 
coverage rates slightly below their nominal value 
($>$85\%). In comparison, as expected (same model as the data generating one), the association 
was correctly estimated with the joint model in all scenarios. We note, however, a slightly reduced coverage rate (87\%) in scenario 12 with the stronger association, larger measurement error and lower number of visits due to lower survival. The empirical distributions of estimates from the MI Cox model and the joint model were consistently wider than those from the other methods, but led in almost all the scenarios to smaller MSEs.

Additional scenarios (Supplementary Figures S3-S8) 
considering quadratic trends of exposure trajectories, 10\% intermittent missing data, correlated random effects, or constant exposure trajectories over time produced findings in line with these results.

\section*{APPLICATION}

\subsection*{The Bordeaux 3C subcohort}
The four methods are illustrated using the 3C cohort (restricted to Bordeaux city), a prospective population-based cohort aiming to study cerebral aging in the elderly, which enrolled in 1999 2104 participants aged 65 years and older, who were randomly selected from the electoral rolls. The participants were interviewed at home by a trained neuropsychologist at baseline and then at 2, 4, 7, 10, 12, 14, and 17 years of follow-up. The diagnosis of dementia was established using a three-stage strategy with final confirmation by an expert committee of external neurologists.

The aim was to assess the association of two time-varying covariates with the risk of dementia: Body Mass Index (BMI) and the Trail Making Test part A (TMT-A), a neuropsychological test evaluating executive functions (score defined as the number of correct moves per minute).
The analytical sample included all the participants from Bordeaux who were free of dementia at baseline, who had a measurement of both BMI and TMT-A at baseline and who had no missing data for the three adjustment covariates: educational level, sex and age at inclusion (flowchart in supplementary Figure S9). The final sample comprised 6264 repeated visits for 1754 subjects among which 434 were diagnosed with dementia during the follow-up. The median number of visits per subject was 4; subjects had an average age of 75.9 years at baseline, with a majority being women (61.5\%) and 40.7\% having completed extended secondary education or higher. Supplementary Figure S10 shows BMI and TMT-A over time for 10 randomly selected subjects. 

For ease of implementation, we treated the diagnosis of dementia at follow-up visits as right-censored rather than interval-censored. We defined dementia as occurring at the midpoint between the last negative diagnosis and the positive diagnosis. Participants without a positive diagnosis were censored at their last negative diagnosis. We did not account for the competing risk of death in these examples. 

\subsection*{Specification of the models}
The target model was a proportional hazards model for time since baseline. The linear predictor included the current level of the time-varying covariate of interest (BMI or TMT-A) and was adjusted for age at baseline, sex and educational level. For TMT-A only, a binary indicator for the first visit was also included in the model to account for the practice effect \cite{vivot}. The baseline risk was left unspecified for LOCF, RC and MI Cox, and parameterized using B-splines for JM (1 interior knot). 

The longitudinal exposure trajectories were flexibly modeled for the RC, MI and JM methods with natural cubic splines on time and with individual random effects on the intercept and on each spline function (three correlated random effects in total). Additionally, this model was adjusted for age at baseline, sex and educational level, including fixed main effects and interactions with the spline functions for time. 

\subsection*{Results}
The estimated association between current BMI and the instantaneous risk of dementia was virtually identical for all methods, with higher BMI associated with a lower concomitant risk of dementia (Figure \ref{fig:appli}, Supplementary Table S2). Log-hazard ratios (log-HR) ranged from -0.046 (95\% CI -0.071,-0.021) to -0.043 (95\% CI -0.069,-0.017) for one unit increase of BMI, suggesting a weak association that is robust across methods.
The estimated association of current executive functioning (TMT-A score) with dementia risk showed much more variation across methods, but all agreed that higher executive functioning was associated with lower risk of dementia (Figure \ref{fig:appli}, Supplementary Table S2). 
The LOCF and RC methods resulted in the smallest estimates with log-HR of -0.071 (95\% CI -0.084,-0.058) and -0.088 (95\% CI -0.105,-0.070) compared to JM and MI-Cox with log-HR of -0.106 (95\% CI -0.129,-0.084) and -0.122 (95\% CI -0.148,-0.097), indicating an attenuated estimation for the two former methods.  

\section*{DISCUSSION}

Although commonly encountered in epidemiological studies, the association of a survival endpoint and time-varying exposures that are infrequently measured and prone to error demands special attention. Acknowledging that joint models require advanced statistical expertise and that existing software is  not applicable in all situations (e.g., limitations in the number of variables and the types of associations), we investigated the performance of simpler approaches. Our simulations confirmed that classical approaches such as LOCF and, to a lesser extent, classical RC usually lead to attenuated associations between exposure and survival, and should therefore be avoided. Although bias to the null has already been noted by others\cite{andersen,prentice1982}, LOCF is still quite commonly used in epidemiology. In contrast to LOCF and RC, the MI technique performs well; note that we still observed a residual bias and low coverage rates for the 95\% confidence intervals in cases of strong association or large measurement error. 

Our work highlights the limitations of the classical regression calibration in longitudinal studies when the event causes informative truncation of the exposure measurements. RC relies on the missing at random assumption for the infrequently observed exposures to correctly retrieve the underlying process. As shown in the simulations, its use should be limited to very specific scenarios where the association is weak and the truncation is mild. Risk-Set Regression Calibration extends the regression calibration technique to the informative truncation context \cite{liao}. However, it is numerically very demanding and requires more advanced expertise in the absence of standard software so we did not cover it in this work. 

In our application, we compared the modeling approaches using two different exposure processes: BMI, which didn't change much over time and which association with dementia was weak (although significant at the 5\% level), and TMT-A, which changed more markedly over time and exhibited a stronger significant association with dementia. While the RC and LOCF showed a pronounced attenuation of the association compared to MI and JM when applied to TMT-A, both RC and LOCF gave roughly the same results as the other methods when applied to BMI. Our simulations indeed suggested that the LOCF and RC may provide correct estimates in very specific cases where the association is weak and the measurement error is small. 
Previous research \cite{arisido, nevo} has also demonstrated that LOCF performs relatively well with a time-varying covariate not associated with the event, and our additional simulation scenarios 38-43 (Figure S8) confirmed this. 

It should be noted that all the results reported here apply to any time-varying exposure, whether endogenous or exogenous, that is measured infrequently, is subject to classical measurement error, and whose measurements cease at the event time. 
For ease of presentation, we limited our simulations to a proportional hazards model with a single exposure having a linear and concomitant relation to the log-hazard. However, we expect our conclusions to extend to any number of time-varying exposures and any form of association provided the form of association has been carefully assessed and is correctly specified. In the applications, we also considered, for illustrative purposes only, the associations of the time-varying exposure with simultaneous dementia risk. This may explain the protective association observed with BMI, which could reflect reverse causation, as BMI changes while the disease progresses in its subclinical stages \cite{wagner2021}.

We focused in this work on LOCF and two-stage approaches of RC and MI. Instead of estimating the true exposure in a first stage as done in RC and MI, Wang et al proposed to correct the LOCF approach \cite{wang-abrahamowicz} by (i) adjusting for the elapsed time since the last measurement to address sparsity and (ii) using the Simulation-Extrapolation (SIMEX, \cite{simex}) method to correct for measurement error. A formal comparison with this correction-based approach is left for future research. 

Given the burdensome estimation of joint models and the required advanced statistical expertise, two-stage techniques offer a promising alternative. By relying on classical tools for survival data, they scale more easily to high-dimensional contexts and complex dependency structures. However, it is crucial to account for the specific characteristics of the longitudinal data (e.g., sparse and irregular visits, measurement error) and the informative truncation of the exposure process in the first stage. Imputation techniques that incorporate information about the event into the first-stage longitudinal model proved to be a promising solution. Following \cite{moreno}, we included the cumulative risk as the event variable that needs to be included in the model when MI is used but alternatives could also be considered.


This comparative study aimed to highlight the consequences of ignoring the infrequent measurements and errors of time-varying exposures in survival analyses, a common issue in epidemiology. Based on the simulations and case studies presented, we  encourage analysts to favor appropriate methods, such as joint models where feasible or other specifically-tailored techniques including carefully designed two-stage approaches. 

\bibliography{timedepvardraft}

\begin{thebibliography}{10}

\bibitem{cox1972}
Cox David~R. Regression models and life-tables  {\it Journal of the Royal
  Statistical Society: Series B (Methodological). } 1972;34:187--202.

\bibitem{prentice1982}
Prentice R.~L.. Covariate measurement errors and parameter estimation in a
  failure time regression model  {\it Biometrika. } 1982;69:331--342.

\bibitem{kalbfleisch}
Kalbfleisch John~D., Schaubel Douglas~E.. Fifty Years of the Cox Model  {\it
  Annual Review of Statistics and Its Application. } 2023;10:1-23.

\bibitem{arisido}
Arisido Maeregu~W., Antolini Laura, Bernasconi Davide~P., Valsecchi Maria~G.,
  Rebora Paola. Joint model robustness compared with the time-varying covariate
  Cox model to evaluate the association between a longitudinal marker and a
  time-to-event endpoint  {\it BMC Medical Research Methodology. } 2019;19.

\bibitem{liao}
Liao Xiaomei, Zucker David~M., Li~Yi, Spiegelman Donna. Survival Analysis with
  Error-Prone Time-Varying Covariates: A Risk Set Calibration Approach  {\it
  Biometrics. } 2011;67:50-58.

\bibitem{tsiatis}
Tsiatis Anastasios~A., Davidian Marie. A Semiparametric Estimator for the
  Proportional Hazards Model with Longitudinal Covariates Measured with Error
  {\it Biometrika. } 2001;88:447--458.

\bibitem{wang}
Wang Ching-Yun, Song Xiao. Semiparametric regression calibration for general
  hazard models in survival analysis with covariate measurement error;
  surprising performance under linear hazard  {\it Biometrics. }
  2021;77:561-572.

\bibitem{andersen}
Andersen Per~Kragh, Liestøl Knut. {Attenuation caused by infrequently updated
  covariates in survival analysis}  {\it Biostatistics. } 2003;4:633-649.

\bibitem{laird1982}
Laird N.~M., Ware J.~H.. Random-effects models for longitudinal data  {\it
  Biometrics. } 1982;38:963--74.

\bibitem{ye2008}
Ye~Wen, Lin Xihong, Taylor Jeremy M~G.. Semiparametric modeling of longitudinal
  measurements and time-to-event data–a two-stage regression calibration
  approach.  {\it Biometrics. } 2008;64:1238--46.

\bibitem{shaw2018}
Shaw Pamela~A., Deffner Veronika, Keogh Ruth~H., et al. Epidemiologic analyses
  with error-prone exposures: review of current practice and recommendations
  {\it Annals of Epidemiology. } 2018;28:821--828.

\bibitem{boe}
Boe Lillian~A, Shaw Pamela~A, Midthune Douglas, et al. Issues in implementing
  regression calibration analyses  {\it American Journal of Epidemiology. }
  2023;192:1406--1414.

\bibitem{geskus}
Geskus Ronald~B. Which individuals make dropout informative?  {\it Statistical
  methods in medical research. } 2014;23:91--106.

\bibitem{rizopoulosbook}
Rizopoulos Dimitris. {\it Joint models for longitudinal and time-to-event data:
  With applications in R}.
\newblock CRC press 2012.

\bibitem{wulfsohn}
Wulfsohn Michael~S, Tsiatis Anastasios~A. A joint model for survival and
  longitudinal data measured with error  {\it Biometrics. } 1997:330--339.

\bibitem{rustand}
Rustand Denis, Niekerk Janet, Rue H{\aa}vard, Tournigand Christophe, Rondeau
  Virginie, Briollais Laurent. Bayesian estimation of two-part joint models for
  a longitudinal semicontinuous biomarker and a terminal event with INLA:
  Interests for cancer clinical trial evaluation  {\it Biometrical Journal. }
  2023;65:2100322.

\bibitem{hickey}
Hickey Graeme~L, Philipson Pete, Jorgensen Andrea, Kolamunnage-Dona Ruwanthi.
  joineRML: a joint model and software package for time-to-event and
  multivariate longitudinal outcomes  {\it BMC medical research methodology. }
  2018;18:1--14.

\bibitem{jmbayes2}
Rizopoulos Dimitris, Papageorgiou Grigorios, Miranda~Afonso P. JMbayes2:
  extended joint models for longitudinal and time-to-event data  {\it R
  package. v. 0.4-5. Available online: https://drizopoulos. github.
  io/JMbayes2/(accessed on 26 July 2022). } 2022.

\bibitem{moreno}
Moreno-Betancur Margarita, Carlin John~B, Brilleman Samuel~L, Tanamas
  Stephanie~K, Peeters Anna, Wolfe Rory. {Survival analysis with time-dependent
  covariates subject to missing data or measurement error: Multiple Imputation
  for Joint Modeling (MIJM)}  {\it Biostatistics. } 2017;19:479-496.

\bibitem{efron1994}
Efron B., Tibshirani R.J.. {\it An Introduction to the Bootstrap}.
\newblock Chapman \& Hall/CRC Monographs on Statistics \& Applied
  ProbabilityTaylor \& Francis 1994.

\bibitem{rubin}
Rubin Donald~B. {\it Multiple imputation for nonresponse in surveys};81.
\newblock John Wiley \& Sons 2004.

\bibitem{philipps}
Philipps Viviane, Hejblum Boris~P, Prague M{\'e}lanie, Commenges Daniel,
  Proust-Lima C{\'e}cile. Robust and efficient optimization using a
  Marquardt-Levenberg algorithm with R package marqLevAlg  {\it arXiv preprint
  arXiv:2009.03840. } 2020.

\bibitem{morris}
Morris Tim~P, White Ian~R, Crowther Michael~J. Using simulation studies to
  evaluate statistical methods  {\it Statistics in medicine. }
  2019;38:2074--2102.

\bibitem{vivot}
Vivot Alexandre, Power Melinda, Glymour M., et al. Jump, Hop, or Skip: Modeling
  Practice Effects in Studies of Determinants of Cognitive Change in Older
  Adults  {\it American Journal of Epidemiology. } 2016;183:kwv212.

\bibitem{nevo}
Nevo Daniel, Hamada Tsuyoshi, Ogino Shuji, Wang Molin. {A novel calibration
  framework for survival analysis when a binary covariate is measured at sparse
  time points}  {\it Biostatistics. } 2018;21:e148-e163.

\bibitem{wagner2021}
Wagner Maude, Grodstein Francine, Leffondre Karen, Samieri* Cécilia,
  Proust-Lima* Cécile. Time-varying associations between an exposure history
  and a subsequent health outcome: a landmark approach to identify critical
  windows  {\it BMC medical research methodology. } 2021;21:266.

\bibitem{wang-abrahamowicz}
Wang Yishu, Beauchamp Marie-Eve, Abrahamowicz Michal. Nonlinear and
  time-dependent effects of sparsely measured continuous time-varying
  covariates in time-to-event analysis  {\it Biometrical Journal. }
  2020;62:492-515.

\bibitem{simex}
Stefanski L.~A., Cook J.~R.. Simulation-Extrapolation: The Measurement Error
  Jackknife  {\it Journal of the American Statistical Association. }
  1995;90:1247-1256.

\end{thebibliography}
\bibliographystyle{ama}

\begin{figure}[h!]
  \includegraphics[width=1.0\textwidth]{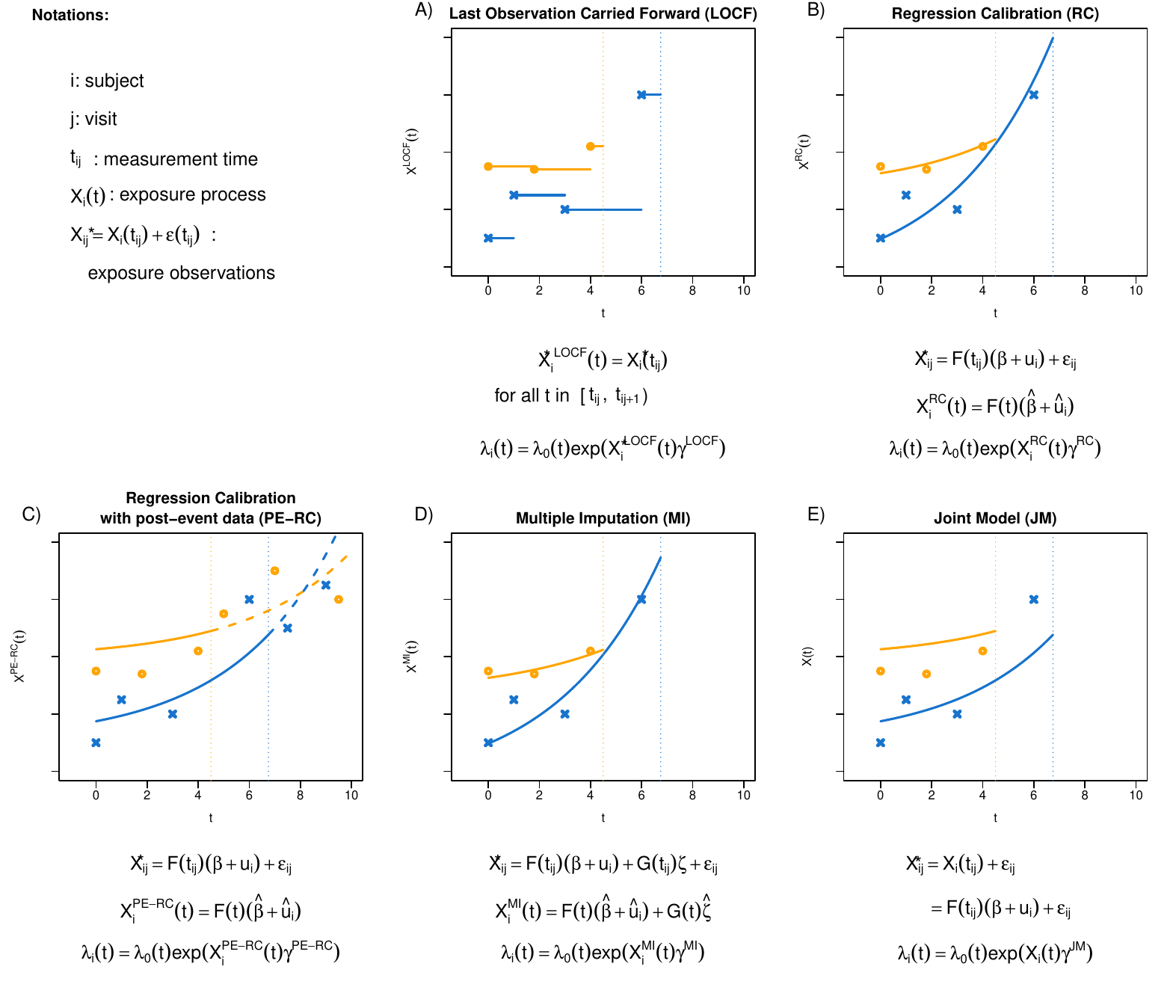}
  \caption{Illustration of five methods dealing with time-varying covariates in survival analyses: last observation carried forward (LOCF), regression calibration (RC), regression calibration with post-event information (PE-RC), multiple imputation (MI), and joint model (JM). The orange points represent the observations of a first fictive subject, the blue crosses the ones of a second subject. Plain lines are the covariate's values considered in each approach. Vertical dotted lines represent the event times. }
  \label{fig:methods}
\end{figure}

\begin{figure}[h!]
  \includegraphics[width=1.0\textwidth]{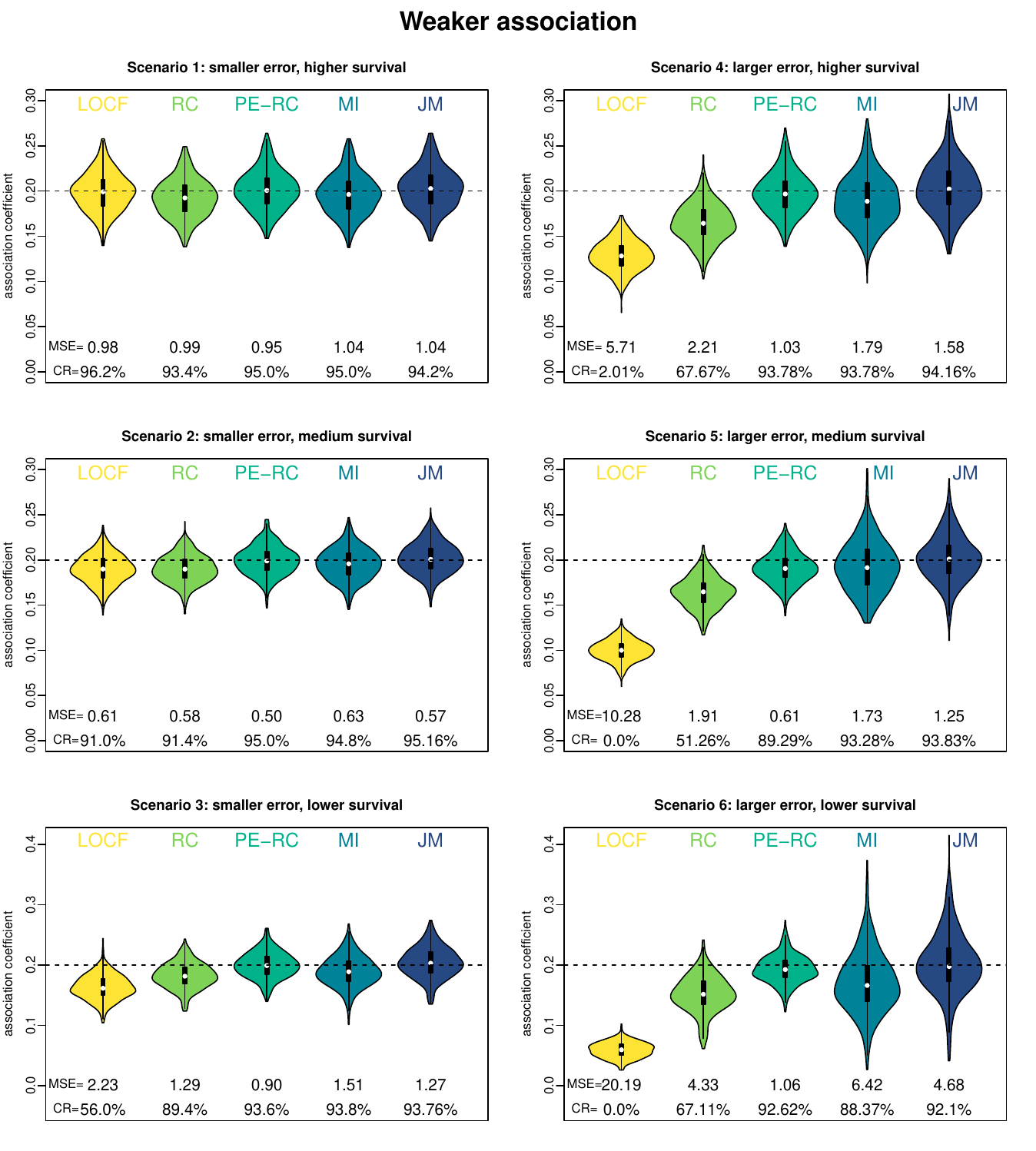}
  \caption{Estimated association coefficient in the survival model over 500 simulation replicates when considering last observation carried forward (LOCF), regression calibration (RC), regression calibration with post-event information (PE-RC), multiple imputation (MI) and joint model (JM) methods. The true association value 0.2 is indicated by a dashed line. Are reported the results for a measurement error magnitude of 1 (left column) and 3 (right column) and different baseline survival (high on the top, medium in the middle and low on the bottom). The coverage rates (CR) of the 95\% confidence intervals and the mean square error (MSE, x1000) are given at the bottom of each panel.}
   \label{fig:sim1}
\end{figure}

\begin{figure}[h!]
 \includegraphics[width=1.0\textwidth]{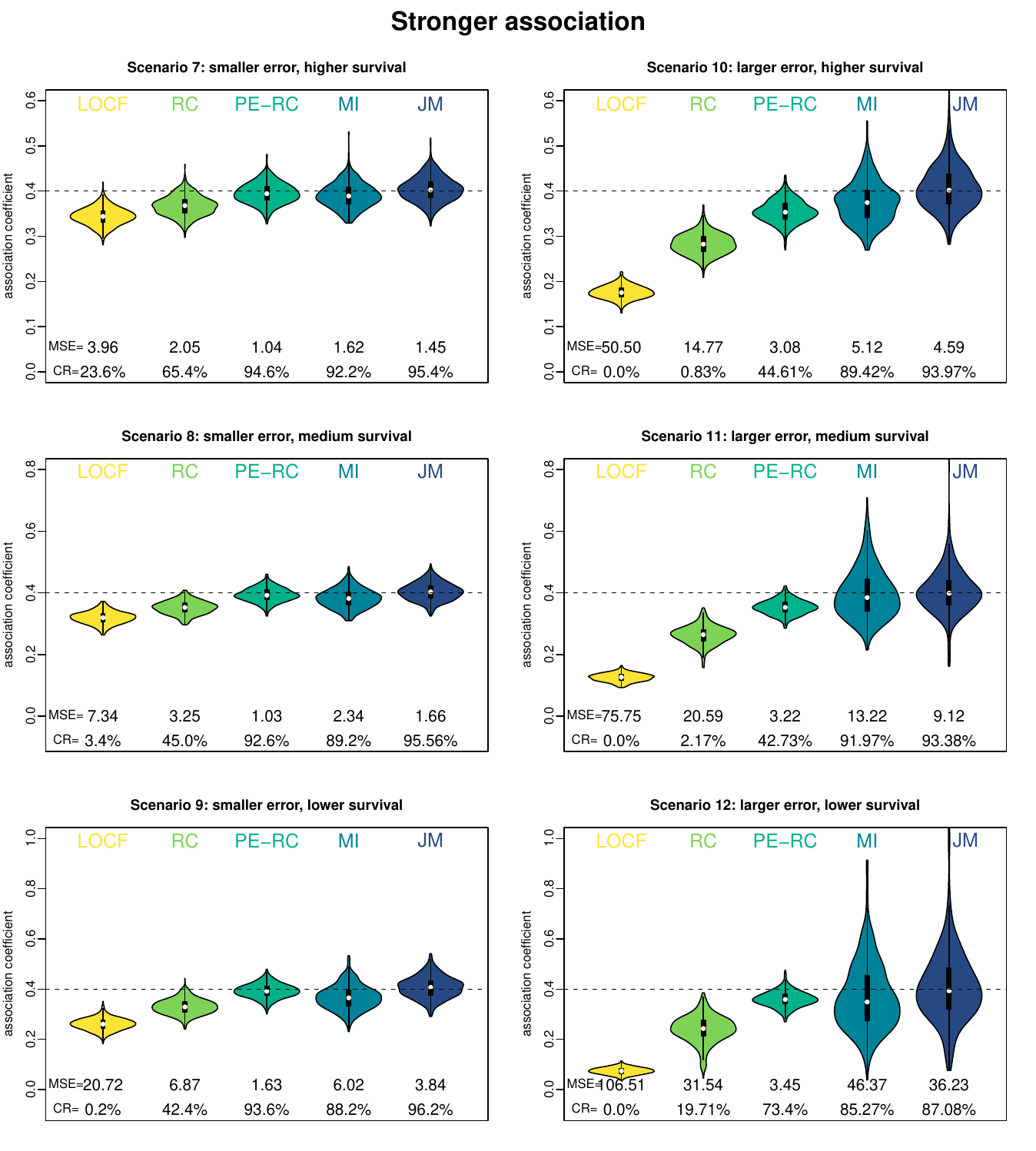}
 \caption{Estimated association coefficient in the survival model over 500 simulation replicates when considering last observation carried forward (LOCF), regression calibration (RC), regression calibration with post-event information (PE-RC), multiple imputation (MI) and joint model (JM) methods. The true association value 0.4 is indicated by a dashed line. Are reported the results for a measurement error magnitude of 1 (left column) and 3 (right column) and different baseline survival (high on the top, medium in the middle and low on the bottom). The coverage rates (CR) of the 95\% confidence intervals and the mean square error (MSE, x1000) are given at the bottom of each panel.}
  \label{fig:sim2}
\end{figure}

\begin{figure}[h!]
 \includegraphics[width=1.0\textwidth]{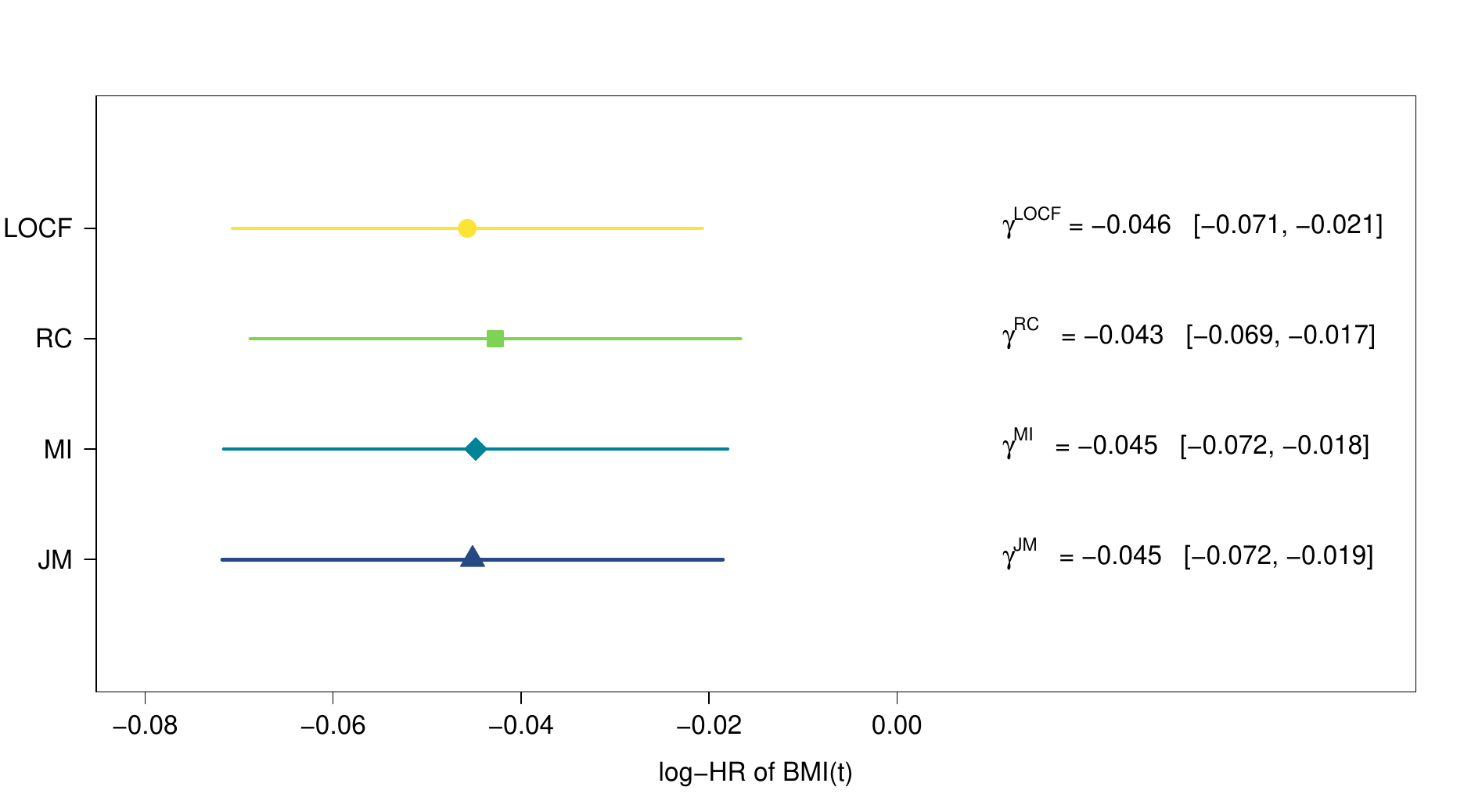}
 
 \includegraphics[width=1.0\textwidth]{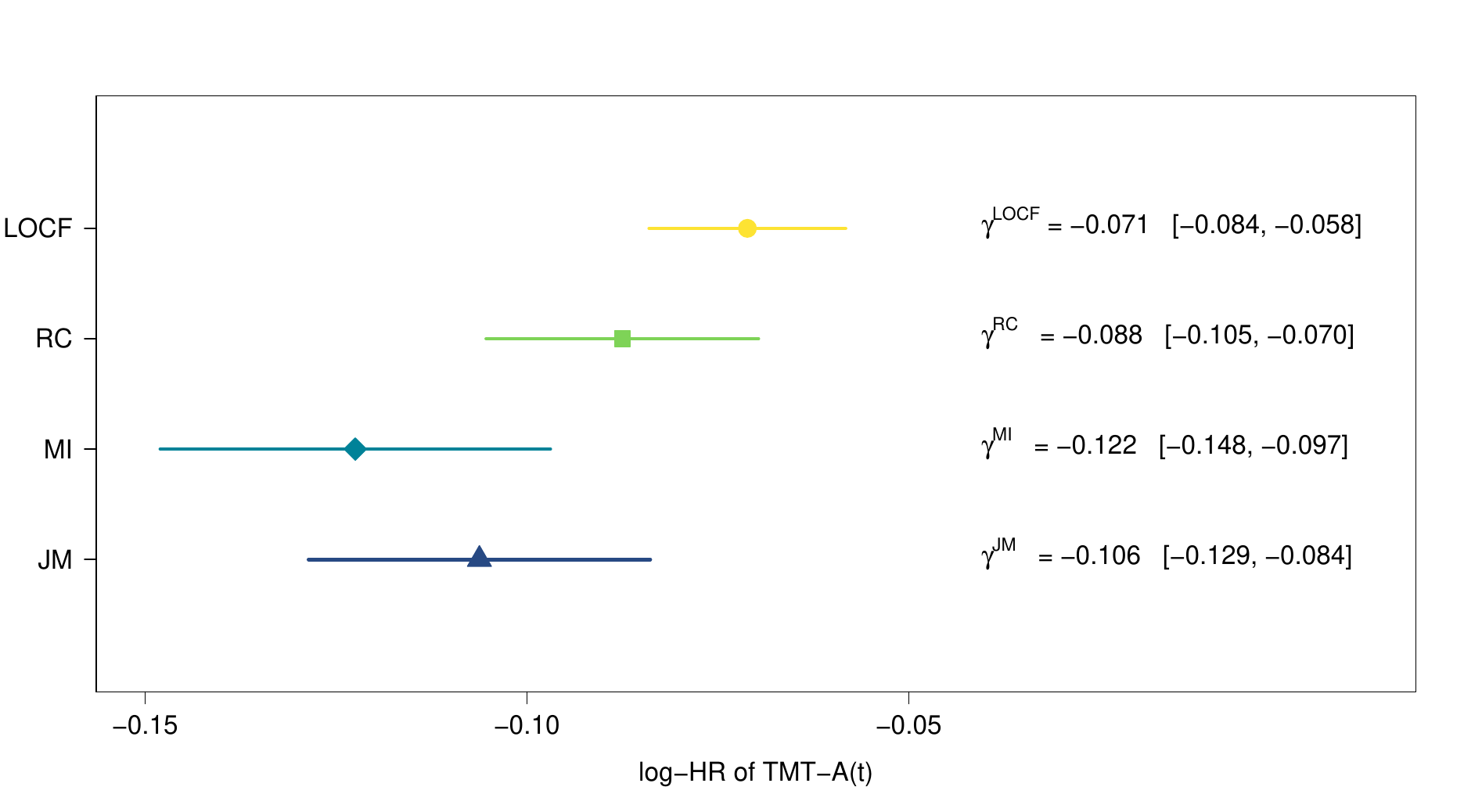}
 \caption{Log-hazard ratios (log-HR) with 95\% conficence intervals (plain lines) between the time-varying BMI (top panel) {and} the TMT-A (bottom panel) and the risk of dementia estimated using {last observation carried forward (LOCF) Cox model, regression calibration (RC) Cox model, multiple imputation (MI) Cox model and joint model (JM)}. Estimates are reported for a 1-unit increase and are adjusted for age at baseline, sex and educational level (Bordeaux 3C {sub}cohort, N=1754).}
  \label{fig:appli}
\end{figure}

\end{document}